\documentclass[preprint,aps,nofootinbib]{revtex4}
\usepackage{}
\usepackage{graphicx}
\usepackage{amsmath}
\usepackage{amsfonts}
\usepackage{amssymb}
\usepackage{color}%
\usepackage{dcolumn}
\providecommand{\U}[1]{\protect\rule{.1in}{.1in}}

\definecolor{lightgray}{rgb}{.7,.7,.7}

\definecolor{red}{rgb}{1,0,0}

\definecolor{blue}{rgb}{0,0,1}

\newcommand{\f}{\begin{equation}}
\newcommand{\ff}{\end{equation}}
\newcommand{\fa}{\begin{eqnarray}}
\newcommand{\ffa}{\end{eqnarray}}

\begin{document}
\title{Fluid/gravity correspondence for massive gravity}
\author{Wen-Jian Pan $^{1}$}
\email{wjpan_zhgkxy@163.com}
\author{Yong-Chang Huang $^{1}$}
\email{ychuang@bjut.edu.cn}
\affiliation{$^1$ Institute of theoretical physics,Beijing University of Technology,Beijing,100124,China}

\begin{abstract}
In this paper, we investigate the fluid/gravity correspondence
in the framework of massive Einstein gravity. Treating the gravitational
mass terms as an effective energy-momentum tensor and utilizing
the Petrov-like boundary condition on a timelike hypersurface,
we find that the perturbation effects of massive gravity in bulk
can be completely governed by the incompressible Navier-Stokes
equation living on the cutoff surface under the near horizon and
nonrelativistic limits. Furthermore, we have concisely computed
the ratio of dynamical viscosity to entropy density for two
massive Einstein gravity theories, and found that they still
saturate the Kovtun-Son-Starinets (KSS) bound.
\end{abstract}

\maketitle

\section {Introduction}
Recently, in the context of the AdS/CFT correspondence \cite{Maldacena:1997re,
Gubser:1998bc,Witten:1998qj,Aharony:1999ti}, the fluid/gravity
duality becomes one of most interesting topics rooted in the observation
by Damour that the gravitational perturbation
behavior of the black hole horizon was very close to those of
a fluid \cite{Damour1979}. As a result, the hydrodynamical behavior of
gravity has been heavily investigated and a great many of works have been made \cite{P-T,Jacobson:1995ab,PSS,KSS,B-L,I-L,
Bhattacharyya:2008kq,EFO,Padmanabhan10rp,Wilsonian,Heemskerk10hk,Faulkner10jy,
Bredberg:2011jq,Compere:2011dx,Cai,Bredberg:2011xw,Niu:2011gu,Matsuo11fk,
Lysov11xx,HL,Huang:2011kj,Zhang:2012uy,Wu:2013kqa,Wu:2013mda,
Ling:2013kua,Cai:2013uye,Cai:2014ywa,Cai:2014sua,Hao:2014xva,Hao:2015zxa,
Pan:2015kia,Hao:2015pal,Wu:2015pzg,Chou:2016qij,Fujisawa:2015riu,
Rebhan:2011vd,Cheng:2014sxa,Cheng:2014qia,Ge:2014aza,
Jain:2014vka,Critelli:2014kra,Jain:2015txa,
Bu:2014sia,Bu:2014ena,Donos:2015gia,Banks:2015wha,
Bu:2015ika,Bu:2015bwa,Bu:2015ame,Blake:2015epa,Blake:2015hxa}. In particular, recent developments
on fluid/gravity duality in the context of AdS/CFT framework have
shed more insightful light on their relationships.

Specifically, in the hydrodynamical limit, the method \cite{Wilsonian,Bredberg:2011jq,Compere:2011dx,Cai,
Bredberg:2011xw,Niu:2011gu,Matsuo11fk} of directly perturbing metric in bulk
was proposed to construct the dual fluid equation living on the timelike
hypersurface, in which the regularity on the horizon and the Dirichlet
boundary condition on the surface were imposed. Furthermore, the authors of \cite{Bredberg:2011jq} have
shown that the perturbation solution of gravity in long
wavelength limit satisfies the Petrov-like boundary condition,
and they have found that in the fluid/gravity duality the near-horizon
expansion was mathematically equivalent to
the long wavelength expansion.

Very remarkably, indeed an alternative approach in \cite{Lysov11xx} was proposed
to construct the fluid/gravity duality in Rindler spacetime via
imposing the Petrov-like condition on the cutoff
surface. It has been shown that, under the near horizon limit
and the nonrelativistic limit, embedding a
hypersurface $\Sigma_c$ into a Rindler spacetime,
the gravitational fluctuation can reduce exactly to
the incompressible Navier-Stokes equation moving on the cutoff surface
with one lower dimension. More specifically, in this strategy,
keeping the induced metric fixed and treating the extrinsic curvature
as a fundamental variable, the Petrov-like condition
on hypersurface exactly reduces the degrees of freedom of
the extrinsic curvature to those of dual fluid. Finally,
with the use of ``momentum constraint'' and ``Hamiltonian constraint,''
those unconstrained variables can give exactly rise to
the incompressible Navier-Strokes equation. Recently,
since the method of constructing dual fluid
by a Petrov-like condition is very elegant and powerful,
it has been further developed
in \cite{HL,Huang:2011kj,Zhang:2012uy,Wu:2013kqa,Wu:2013mda,
Ling:2013kua,Cai:2013uye,Cai:2014ywa,Cai:2014sua,Hao:2014xva,
Hao:2015zxa,Pan:2015kia,Hao:2015pal,Wu:2015pzg,Chou:2016qij}.
In particular, it has been even applied to effectively describe
the hydrodynamical behavior of gravity with spatially curved
spacetime \cite{HL,Huang:2011kj,Zhang:2012uy,Wu:2013kqa,Wu:2013mda}.

On the other hand, in the context of AdS/CFT correspondence, there have
been many applications for massive gravity theories
\cite{Vegh:2013sk,Blake:2013bqa,Andrade:2013gsa,Amoretti:2014mma,Davison:2013jba,
Zeng:2014uoa,Baggioli:2014roa,Zhou:2015dha,Sadeghi:2015vaa,Alberte:2016xja,Hartnoll:2016tri}
that arose from the dRGT theory \cite{deRham:2010kj}. Because of translation invariance violated in these theories,
this provides a momentum dissipation effect
which is equivalent to the one of the lattice model or momentum
relaxation to describe and understand many physical quantities
by holographic dual in the condensed matter theory (CMT),
such as the resistivity in CMT, finite direct current (DC)
conductivity, thermoelectric conductivity, Hall angle, and so on.

Very recently, there have been some discussions on the dual hydrodynamical
behavior in massive gravity theory, where the transport coefficients
have been evaluated in literature \cite{Davison:2013jba,Sadeghi:2015vaa,Alberte:2016xja,Hartnoll:2016tri}.
Nevertheless, one still knows little information about
the connection between the massive gravity and the dual
fluid dynamics. Naturally, there is an interesting problem to ask
whether such violating behavior in massive gravity influences its dual fluid dynamical behavior or not.
A complete resolution to this problem is still absent so far.

In this paper, our goal is to establish the dual relationship
between the pure massive Einstein gravity and the nonlinear
Navier-Stokes equation in the framework with the Petrov-like
boundary constraint. It turns out that the near horizon perturbation behavior
of the massive gravity can be governed completely by the incompressible
Naiver-Stokes equation residing on embedded hypersurface with one lower dimension
through imposing a Petrov-like boundary condition on the cutoff
surface in the nonrelativistic limit and the near horizon limit.
Furthermore, in such boundary condition, the kinematic viscosity
value in the dual fluid equations implies that the ratio of $\frac{\eta}{s}$
still saturates the Kovtun-Son-Starinets (KSS) bound \cite{Kovtun:2004de}.

The rest of our paper is organized as follows. In Sec. II
we briefly recall some important ingredients to study
the fluid/gravity duality in the Petrov-like framework. In Secs. III and IV
imposing Petrov-like boundary condition in the nonrelativistic limit and the near horizon limit,
we will in detail derive the corresponding incompressible
Navier-Strokes equations moving on a spatially
flat hypersurface from the pure massive gravity
theories, respectively, and discuss the behavior of $\frac{\eta}{s}$
in these models. In the last section we will
give our summary and some discussions.

\section{some useful formulas}
Before we study the dual relation between the massive gravity
and the nonlinear Navier-Stokes equation, we would like to recall
some important ingredients, pioneered in \cite{Lysov11xx,HL,Huang:2011kj},
as the roles of bridge on the fluid/gravity duality. First of all,
we naturally require that a p+2 dimensional manifold
with a cosmological constant $\Lambda$ and an energy momentum
tensor $T_{\mu\nu}$ should be
described by the Einstein theory:
\begin{eqnarray}\label{EE}
G_{\mu\nu}=-{\Lambda}g_{\mu\nu}+T_{\mu\nu},\quad{\mu},{\nu}=0,{\ldots},p+1,
\end{eqnarray}
where $g_{\mu\nu}$ is a metric of the $p+2$ dimensional spacetime.
Second, in order to extract the hydrodynamical behavior dual to the gravity
in the $p+2$ dimensional bulk space, we need to embed
a timelike hypersurface $\Sigma_c$ with an induced metric $\gamma_{ab}$
into this bulk space, where its extrinsic curvature
$K_{ab}$ should satisfy the momentum constraint
 \begin{eqnarray}
 D^a(K_{ab}-{\gamma_{ab}}K)=T_{\mu b}n^{\mu},\label{mc}
 \end{eqnarray}
 as well as the Hamiltonian constraint
 \begin{eqnarray}\label{HC1}
{^{p+1}R}+{K_{ab}}{K^{ab}}-{K^2}-{2\Lambda} = -2T_{\mu\nu}n^{\mu}n^{\nu},
\end{eqnarray}
where $D_a$ is compatible with the induced metric on $\Sigma_c$,
namely $D_{a}{\gamma_{bc}}=0$, $K$ is the trace of extrinsic
curvature and $n^{\mu}$ is the unit normal to $\Sigma_c$. It is worth noting that
the Hamiltonian constraint becomes an equation of state relating the energy density
with the pressure in dual fluid on cutoff surface, while
the momentum constraint gives rise to the dynamical equation of the dual fluid.

For the method imposing Petrov-like condition on this cutoff surface,
its framework has been specifically constructed
in previous literature \cite{Lysov11xx,HL,Huang:2011kj}.
And the Petrov-like condition on the cutoff surface $\Sigma_c$ is defined as
\begin{eqnarray} \label{petrov}
C_{(\ell)i(\ell)j}=\ell^{\mu}{m_i}^{\nu}\ell^{\alpha}{m_j}^{\beta}C_{\mu\nu\alpha\beta}=0
\end{eqnarray}
 where the corresponding $p+2$ Newman-Penrose-like
vector fields in Eq.(\ref{petrov}) satisfy the following relations
\begin{eqnarray}
\ell^2=k^2=0,\ \ (k,\ell)=1,\ \ (k,m_i)=(\ell,m_i)=0,\ \
(m^i,m_j)={\delta^i}_j.
\end{eqnarray}
The realization from Einstein gravity to the Navier-Stokes equation
by the method of imposing the Petrov-like condition can be understood by
counting the degrees of freedom of the extrinsic curvature $K_{ab}$ on hypersurface. The Petrov-like
boundary condition provides $\frac{p(p+1)}{2}-1$ constraints on
the extrinsic curvature to reduce its $\frac{(p+1)(p+2)}{2}$ independent
components to $p+2$ unconstrained variables which are identified
as the energy density, pressure and $p$ velocity fields in dual fluid.
In this sense, the Petrov-like condition plays a holographic role on the hydrodynamical
behavior of gravity. In the following two sections, based on the framework,
we will take two pure massive Einstein gravity models as examples and in details construct
the fluid/gravity duality in these models.

\section{dual fluid in the simplest massive gravity}
In this section, taking the simplest massive theory as an example,
we demonstrate how it gives rise to the Navier-Stokes equation
in the Petrov-like framework, which can be useful for us to
further understand the fluid/gravity duality.
To do this, we consider a 4-dimensional massive Einstein-Hilbert
action with a surface term and a massive term $m^2\alpha_1 tr\mathcal{K}$ \cite{Vegh:2013sk,Hartnoll:2016tri},
which is written as
\begin{equation}\label{s1}
S=\int d^4x\sqrt{-g}[\frac{1}{2\kappa^2}(R+\frac{6}{L^2})+\frac{m^2}{\kappa^2}\alpha_1 tr\mathcal{K}]-\frac{1}{\kappa^2}\int d^3x\sqrt{-\gamma}K
\end{equation}
where $\alpha_1$ is a negative constant, and $\mathcal{{K^{\mu}}_{\nu}}=\sqrt{g^{\mu\alpha}f_{\alpha\nu}}$.
This massive gravity model couples the metric tensor $g_{\mu\nu}$
to a fixed reference metric $f_{\mu\nu} $, giving a mass $m$ to $g_{\mu\nu}$
and breaking diffeomorphism invariance. The reference metric is chosen to be
$f_{\mu\nu}=diag(0,0,1,1)$, which breaks
the translational symmetry along two spatial directions. And $K$ is the trace
of the extrinsic curvature on a timelike hypersurface $\Sigma_c$ with
the induced metric $\gamma_{ab}$. Here and henceforth,
we will set the cosmological scale as $L=1$.
One can obtain the equation of motion for the massive gravity
by varying the above action with respect to the metric,
 \begin{equation}\label{EOM}
G_{\mu\nu}=3g_{\mu\nu}-m^2\alpha_1(\mathcal{K_{\mu\nu}}-tr\mathcal{K}g_{\mu\nu}).
\end{equation}
Here we have used $\Lambda=-3$ and take $\kappa^2\equiv8 \pi G=1$.
Comparing Eqs.(\ref{EE}) and (\ref{EOM}), on mathematical form,
we can take the mass term on the right side of Eq.(\ref{EOM})
to define an effective energy-momentum tensor,
\begin{equation}
T_{\mu\nu}=-m^2\alpha_1(\mathcal{K_{\mu\nu}}-g_{\mu\nu}tr\mathcal{K}).
\end{equation}
Note that this tensor of the massive gravity theory
differing from the usual energy-momentum tensor
is completely determined by the bulk metric.

To investigate the dual hydrodynamical behavior in this model, we
assume that the background geometry has the following form in the
Eddington-Finkelstin coordinates,
\begin{equation}\label{gmn}
ds^2_4=-r^2f(r)dt^2+2dtdr+r^2\delta_{ij}dx^idx^j, \quad{i},{j}=1,2.
\end{equation}
 According to this metric ansatz, we can solve the equation of motion in (\ref{EOM})
and find that the action (\ref{s1}) admits the background solution, \footnote{ When
$\alpha_1\to0$, the background geometry reduces exactly to the black brane solution in \cite{Huang:2011kj}.}
 which turns out to be
\begin{equation}
f(r)=1-\frac{r^3_h}{r^3}+\frac{m^2\alpha_1}{r}(1-\frac{r^2_h}{r^2}).
\end{equation}
Here we have denoted the position of the black brane horizon as $r_h$ with
$f(r_h)=0$. Since the dual fluid moves on the one less dimensional timelike hypersurface,
now we must embed such a surface into the bulk space by localizing
the position of the hypersurface, namely $r=r_c$, outside the horizon. If we define a function
as $U(r)\equiv r^2f(r)$, then the induced metric on the cufoff surface
can be expressed as
\begin{eqnarray}
ds^2_3=-U(r_c)dt^2+r^2_c\delta_{ij}dx^idx^j\nonumber\\
=-(dx^{0})^2+r^2_c\delta_{ij}dx^idx^j
\end{eqnarray}
where $x^0=\sqrt{U(r_c)}t\equiv\sqrt{U_c}t$. It is very evident that
the embedded hypersurface is intrinsically flat. To present the nonrelativistic
hydrodynamical behavior of the massive gravity model,
we need to introduce a parameter $\lambda$ to redefine a new time coordinate
with $\tau=\lambda x^0$, such that the induced metric can be rewritten as
\begin{eqnarray}
ds^2_3=-\frac{d\tau^2}{\lambda^2}+r^2_c\delta_{ij}dx^idx^j.
\end{eqnarray}
Moreover, we also identify the parameter $\lambda$ with the location
of the hypersurface by setting $r_c-r_h={\beta}^2\lambda^2$, such that
taking $\lambda\rightarrow 0$ means that the near horizon limit
is also achieved. Here we would like to emphasize that when $\lambda\rightarrow 0$,
the mean curvature of hypersurface diverges, which is viewed as
the large mean curvature in \cite{Lysov11xx}. This means that the hypersurface
$\Sigma_c$ is highly accelerated. So far, the parameter $\lambda$ has linked
the nonrelativistic limit and the near horizon limit; hence taking
$\lambda\rightarrow 0$ implies that both the near horizon limit
and the nonrelativistic limit are implemented simultaneously.
The behavior of simultaneously taking these limits is a key
step to derive the constitutive relation in the dual fluid.

Next we are going to consider the behaviors of the extrinsic geometry.
In the coordinate $(\tau, x^i)$, the background components of
extrinsic curvature can be explicitly defined on the timelike
hypersurface $\Sigma_{c}$,
\begin{eqnarray}\label{KB}
{K^{\tau(B)}}_{\tau}&=&\frac{U_c^{\prime}}{2\sqrt{U_c}}\nonumber\\
{K^{\tau(B)}}_i&=&0\nonumber\\
{K^{i(B)}}_j&=&\frac{\sqrt{U_c}}{r_c}{\delta^i}_j\nonumber\\
K^{(B)}&=&\frac{U_c^{\prime}}{2\sqrt{U_c}}+\frac{2\sqrt{U_c}}{r_c}
\end{eqnarray}
where the prime denotes derivative with respect to $r$. In order to
conveniently extract the information of the dual fluid, we need to
define the Brown-York tensors in terms of the extrinsic curvatures \cite{Brown:1992br},
\begin{eqnarray}\label{BYT}
{t^a}_b={\delta^a}_bK-{K^a}_b.
\end{eqnarray}
Here we would like to stress that the mass term in (\ref{s1}) for the massive
Einstein gravity does not change the Brown-York tensor on
the hypersurface at the near horizon. Thus, in this case,
the tensor in (\ref{BYT}) is still valid.
With Eqs.(\ref{KB}) and (\ref{BYT}), considering
the perturbation of Brown-York tensor and the near horizon
expansion of its background part in powers of $\lambda$,
we come to the following relations:
 \begin{eqnarray}\label{CBYT}
{t^{\tau}}_{\tau}&=&\frac{2\sqrt{U^{\prime}_h}}{r_h}\beta\lambda+\lambda{t^{\tau(1)}}_{\tau}+ \ldots \nonumber\\
{t^{\tau}}_i&=&0+\lambda{t^{\tau(1)}}_i+ \ldots \nonumber\\
{t^i}_j&=&(\frac{\sqrt{U^{\prime}_h}}{2\beta\lambda}+\frac{3U^{\prime\prime}_h}{8\sqrt{U^{\prime}_h}}\beta\lambda
+\frac{\sqrt{U^{\prime}_h}}{r_h}\beta\lambda){\delta^i}_j+\lambda{t^{i(1)}}_j+ \ldots \nonumber\\
t&=&2(\frac{\sqrt{U^{\prime}_h}}{2\beta\lambda}+\frac{3U^{\prime\prime}_h}{8\sqrt{U^{\prime}_h}}\beta\lambda
+\frac{2\sqrt{U^{\prime}_h}}{r_h}\beta\lambda)+\lambda t^{(1)}+ \ldots
\end{eqnarray}
So far, we have the behaviors of the Brown-York tensor
in the near horizon and nonrelativistic limits. In the following,
with the relations in (\ref{CBYT}), we wish to find out
the dual fluid relation to the massive gravity theory in these limits.
To do this, we first evaluate the Hamiltonian constraint on
the hypersurface in order to give out the equation of state
in the dual fluid. Specifically, we represent this
constraint in (\ref{HC1}) in terms of the Brown-York tensors,
which turns out to be
\begin{equation}\label{tt}
{t^{\tau}}_{\tau}{t^{\tau}}_{\tau}-\frac{2}{\lambda^2}\gamma^{ij}{t^{\tau}}_i{t^{\tau}}_j-2(\frac{t}{2})^2+{t^j}_i{t^i}_j+6=-\frac{4\alpha_1}{r_c}m^2.
\end{equation}
Here we have used the relevant results (\ref{ttt}) and (\ref{ttr}) in Appendix A.
With the relations in (\ref{CBYT}), Eq.(\ref{tt})
immediately implies that the leading term at the order of $1\over\lambda^2$
automatically vanishes, and the first nontrivial term at the order of $\lambda^0$
has the following form
\begin{equation}\label{ttt1}
{t^{\tau(1)}}_{\tau}=-2\gamma^{(0)ij}{t^{\tau(1)}}_i{t^{\tau(1)}}_j.
\end{equation}
This is just the equation of state in dual fluid in the near horizon
limit and nonrelativistic limit. Note that $i$, $j$ indices are raised
and lowered with $\gamma^{(0)}_{ij}\equiv\gamma(r_h)_{ij}$.

Now we turn to the Petrov-like boundary condition. Substituting Eq.(\ref{BYT}) into Eq.(\ref{petrov}),
this boundary condition can be rewritten as
\begin{eqnarray}\label{Clilj2}
\frac{2}{\lambda^2}\gamma^{ki}{t^{\tau}}_i{t^{\tau}}_j+(\frac{t}{2})^2{\delta^k}_j-{t^{\tau}}_{\tau}(\frac{t}{2}){\delta^k}_j+{t^{\tau}}_{\tau}{t^k}_j
+2\lambda\partial_{\tau}(\frac{t}{2}{\delta^k}_j-{t^k}_j)\nonumber\\
-\frac{\gamma^{ki}}{\lambda}(\partial_i{t^{\tau}}_j+\partial_j{t^{\tau}}_i)-{t^k}_n{t^n}_j+{B^k}_j=0,
\end{eqnarray}
where
\begin{eqnarray}
{B^k}_j=-\frac{1}{2}{\delta^k}_j(T_{00}-2T_{0\alpha}n^{\alpha}+T_{\alpha\beta}n^{\alpha}n^{\beta})+\gamma^{ki}{\gamma_i}^{\alpha}{\gamma_j}^{\beta}R_{\alpha\beta}.
\end{eqnarray}
After performing a detailed calculation in Appendix A,
the first term of ${B^k}_j$ at the order of $\lambda^0$ turns out to be,
\begin{eqnarray}\label{bc}
{B^{k(0)}}_j=-(\frac{2\alpha_1}{r_h}m^2+3){\delta^k}_j.
\end{eqnarray}
Putting Eqs.(\ref{CBYT}) and (\ref{bc}) into (\ref{Clilj2}), one easily
checks that the leading term at the order of $1\over\lambda^2$ automatically
meets the Petrov-like condition
 \begin{equation}
\frac{U^{\prime}_h}{4\beta^2\lambda^2}{\delta^k}_j-\frac{U^{\prime}_h}{4\beta^2\lambda^2}{\delta^k}_j=0,
\end{equation}
and the first nontrivial constitutive relation can be encountered at the order of $\lambda^0$,
\begin{eqnarray}\label{tkj}
{t^{k(1)}}_j=2\gamma^{(0)ki}{t^{\tau(1)}}_i{t^{\tau(1)}}_j-\gamma^{(0)ki}(\partial_i{t^{\tau(1)}}_j+\partial_j{t^{\tau(1)}}_i)
+\frac{t^{(1)}}{2}{\delta^k}_j.
\end{eqnarray}
For the relations (\ref{ttt1}) and (\ref{tkj}) we have
set $\frac{\sqrt{U^{\prime}_h}}{\beta}=1$. Until now
we have obtained the nontrivial relations from the
Petrov-like boundary condition as well.

Finally, we come to the the momentum constraint and
represent the momentum constraint (\ref{mc})
in terms of the Brown-York tensors in a compact form
\begin{equation}
\partial_a{t^a}_b=-T_{b\mu}n^{\mu}\label{mc2}.
\end{equation}

As a result, using the data in both Appendix A and Appendix C,
when $b=\tau$, the momentum constraint with $1\over\lambda$ in the dual holographic fluid
would turn out to be the incompressible condition
\begin{equation}
\partial_it^{\tau i(1)}=0.
\end{equation}
When $b=j$, the momentum constraint at the order of $\lambda^1$ gives rise to
\begin{eqnarray}\label{NS}
\partial_\tau {\upsilon_j} + \upsilon^k\partial_k \upsilon_j -
 \partial^2 \upsilon_j + \partial_jP = 0.
\end{eqnarray}
This is precisely the incompressible Navier-Stokes equation
moving on the cutoff surface. Here we have identified
${t^{\tau(1)}}_i$, and $t^{(1)}$ with the velocity fields
and pressure of dual fluid, respectively. They exactly
have the following forms,
 \begin{eqnarray}
{t^{\tau(1)}}_i=\frac{1}{2}\upsilon_i, \qquad t^{(1)}=P.
\end{eqnarray}
The above Eq.(\ref{NS}) implies that the kinematic shear viscosity is $\nu=1$.
In particular, the ratio of dynamical viscosity to entropy density is
\begin{equation}
\frac{\eta}{s}=\frac{\nu\rho}{s}={\frac{1}{2}\over\frac{1}{4G}}=2G=\frac{1}{4\pi}.
\end{equation}
This directly shows that the ratio of dynamical viscosity to
entropy density still satisfies the KSS bound\cite{Kovtun:2004de}.
Here we have used $8\pi G=1$, and the entropy density $s=\frac{1}{4G}$ in \cite{Cai}.
There are two interesting results in this model. One is that in spite of taking the gravitational
mass term into account, the dual fluid is still the standard incompressible Navier-Stokes equation
without the external force term in the near horizon limit. According to the result in Appendix C,
the perturbation behavior of the gravitational mass term becomes a high order effect
in the holographic dual. The other is that the rate of $\frac{\eta}{s}$
does not violate the KSS bound. Our result for the rate of $\frac{\eta}{s}$
is more like that considered in \cite{Sadeghi:2015vaa,Donos:2015gia,Banks:2015wha},
where it is always $\frac{1}{4\pi}$.
\section{The dual fluid in the general massive gravity}
In this section, we would like to consider its gravitational fluid
behavior for a more general pure massive Einstein gravity theory in a parallel way
in order to compare with those in the last model.
Now let us write out the four dimension action with
a surface term and general gravitational mass
terms for the model \cite{Vegh:2013sk},
\begin{eqnarray}\label{S2}
S=\int d^4x\sqrt{-g}[\frac{1}{2\kappa^2}(R-2\Lambda)+\frac{m^2}{\kappa^2}(\alpha_1u_1+\alpha_2u_2)]-\frac{1}{\kappa^2}\int d^3x\sqrt{-\gamma}K.
\end{eqnarray}
where
\begin{eqnarray}
u_1&=&tr\mathcal{K},\\
u_2&=&(tr\mathcal{K})^2-tr(\mathcal{K}^2),
\end{eqnarray}
and $\alpha_1$, $\alpha_2$ are negative constants.
Varying the action in (\ref{S2}) with respect to metric
field $g_{\mu\nu}$, the equations of motion turns out to be
\begin{eqnarray}
R_{\mu\nu}-{1\over2}Rg_{\mu\nu}-3g_{\mu\nu}=-m^2\alpha_1(\mathcal{K_{\mu\nu}}-tr\mathcal{K}g_{\mu\nu})
-m^2\alpha_2[2(tr\mathcal{K})\mathcal{K}_{\mu\nu}-2{\mathcal{K}_{\mu}}^{\alpha}\mathcal{K}_{\alpha\nu}]\nonumber\\
+m^2\alpha_2 g_{\mu\nu}[((tr\mathcal{K})^2-tr(\mathcal{K}^2))].
\end{eqnarray}
The corresponding effective energy momentum tensor can be defined by
\begin{eqnarray}\label{emt}
T_{\mu\nu}=-m^2\alpha_1(\mathcal{K_{\mu\nu}}-tr\mathcal{K}g_{\mu\nu})
-m^2\alpha_2[2(tr\mathcal{K})\mathcal{K}_{\mu\nu}-2{\mathcal{K}_{\mu}}^{\alpha}\mathcal{K}_{\alpha\nu}
-g_{\mu\nu}((tr\mathcal{K})^2-tr(\mathcal{K}^2))].
\end{eqnarray}
And we assume that the gravitational solution
which follows from (\ref{S2}) still has a following form:
\begin{equation}
ds^2_4=-U(r)dt^2+2dtdr+r^2\delta_{ij}dx^idx^j, \quad{i},{j}=1,2,
\end{equation}
where $U(r)=r^2f(r)$. The solution is determined via solving the equations of motion,
\begin{eqnarray}
U(r)&=&r^2[1-\frac{r^3_h}{r^3}+m^2(\frac{\alpha_1}{r}+\frac{2\alpha_2}{r^2})-m^2\frac{r^3_h}{r^3}(\frac{\alpha_1}{r_h}+\frac{2\alpha_2}{r^2_h})].
\end{eqnarray}
Similarly, To construct the fluid/gravity duality,
we still need to introduce the cutoff surface
with the induced metric,
\begin{eqnarray}
ds^2_3&=&-U(r_c)dt^2+r^2_c\delta_{ij}dx^idx^j\nonumber\\
&=&-(dx^{0})^2+r^2_c\delta_{ij}dx^idx^j\nonumber\\
&=&-\frac{d\tau^2}{\lambda^2}+r^2_c\delta_{ij}dx^idx^j.
\end{eqnarray}
Then, in the coordinate $(\tau,x^i)$,
we still directly consider the perturbation of the Brown-York tensor, and
repeat the previous procedures step by step.
It is not difficult to obtain the data of
the Hamiltonian constraint,
\begin{equation}
{t^{\tau(1)}}_{\tau}=-2\gamma^{(0)ij}{t^{\tau(1)}}_i{t^{\tau(1)}}_j,
\end{equation}
as well as one of Petrov-like boundary condition,
\begin{eqnarray}\label{tkj2}
{t^{k(1)}}_j=2\gamma^{(0)ki}{t^{\tau(1)}}_i{t^{\tau(1)}}_j-\gamma^{(0)ki}(\partial_i{t^{\tau(1)}}_j+\partial_j{t^{\tau(1)}}_i)
+\frac{t^{(1)}}{2}{\delta^k}_j.
\end{eqnarray}
Here we have used the corresponding result ${B^k}_j$ at the order of $\lambda^0$ to calculate out
the above relation (\ref{tkj2}). For the detailed calculation of ${B^k}_j$,
we will present it in Appendix B. Finally we come to
the momentum constraint which causes the incompressible condition
\begin{equation}
\partial_it^{\tau i(1)}=0,
\end{equation}
and standard Navier-Stokes equation
 \begin{eqnarray}
\partial_\tau {\upsilon_j} + \upsilon^k\partial_k \upsilon_j -
 \partial^2 \upsilon_j + \partial_jP = 0.
\end{eqnarray}
Again, the identification relations for the velocity
and pressure of dual fluid are the same as the one in the previous section.
It is easy to find that the same result for the ratio of
dynamical viscosity to entropy density can be still obtained.
\section{Summary and Discussion}
In this paper, motivated by the previous works \cite{Lysov11xx,HL,Huang:2011kj},
we have further shown that treating the mass terms of
the gravitational field as the effective energy-momentum tensor,
the near horizon dynamical behavior of the massive Einstein theories
can be governed by the incompressible Navier-Stokes equation
on the boundary with one lower dimension via employing the
Petrov-like boundary condition on such hypersurface in the
nonrelativistic limit as well as in the near horizon limit.
In each explicit construction, keeping the induced metric
fixed and directly perturbing Brown-York tensors, the Petrov-like
boundary condition in these limits maps the perturbations into
a dual constitutive relation on the embedded hypersurface.
Finally, with the use of the momentum constraint relation,
the dual incompressible Navier-Stokes equation can be derived
successfully from the constitutive relation. Furthermore,
our calculation has shown that, for each model, the perturbation
effect of the massive term becomes a high order behavior in a dual fluid equation. In our setups,
we have concisely computed the ratio of dynamical viscosity to
entropy density for the two massive Einstein gravity theories,
and found that they still saturate KSS bound. In our cases,
The results for the ratio of $\frac{\eta}{s}$ are consistent with
that studied in \cite{Sadeghi:2015vaa}, where the rate is always
equal to $\frac{1}{4\pi}$ when the massive Gauss-Bonnet gravity
reduces to the massive Einstein gravity.

In the simplest massive gravity model, the behavior of $\frac{\eta}{s}$
in our paper is different from the one in literature \cite{Hartnoll:2016tri}
where the KSS bound was violated. For this phenomenon,
there may be a reasonable interpretation that, in the holographic
sense, the different boundary conditions may be responsible
for the different dual field theories. The Petrov-like
boundary condition on the hypersurface in our manuscript
greatly distinguishes from the boundary conditions
in literature \cite{Hartnoll:2016tri}, where the boundary conditions
were the Dirichlet boundary condition and regularity
on the horizon. As a result, there is a different behavior between the ratio of $\frac{\eta}{s}$
arising from our manuscript and the one in the recent literature.
For this argument, it has been shown that the various
boundary conditions indeed give rise to the different
behaviors of the dual field theory \cite{Ling:2013kua,Matsuo:2009pg}.
Nevertheless, a deep understanding of their relationships
between these distinguishing results is still lacking.

There are some interesting challenges: under the nonrelativistic
long-wavelength limit, demanding the Dirichlet boundary condition
and regularity on horizon or the Petrov-like boundary condition, respectively,
one can, in principle, derive the corresponding hydrodynamical behaviors of massive gravity
and transport coefficients on the finite cutoff surface. Then
these results may provide some important hints to bridge and understand
the relationship between the ratio of
$\frac{\eta}{s}$ in our paper and the one in literature \cite{Hartnoll:2016tri}.
The investigation is under progress.

\begin{acknowledgments}
 Wen-Jian Pan would like to thank Yi Ling, Yu Tian, Xiao-Ning Wu,
 Zhen-Hua Zhou, Jian-Pin Wu, Chao Wu, Yun-Long Zhang and Zhuo-Yu Xuan
 for useful discussions. We are also grateful to the anonymous referee for helpful suggestions.
 This work is partly supported by the National Natural Science
 Foundation of China (No. 11275017 and No. 11173028).
\end{acknowledgments}

\section*{Appendix}
\subsection{The energy-momentum tensor in the simplest massive gravity}
In this appendix, we provide the detailed calculation to determine
the effective energy-momentum tensor and the physical quantities ${B^k}_j$
for the simplest massive Einstein gravity system. From the definition of $\mathcal{{K^{\mu}}_{\nu}}$,
it is not difficult to get all the nonzero components of this matric,
\begin{equation}
{\mathcal{K}^i}_j=\frac{1}{r}{\delta^i}_j.\label{sij}
\end{equation}
Then, using the metric in (\ref{gmn}), we can find out the following relations:
\begin{eqnarray}
\mathcal{K}_{ij}&=&r\delta_{ij},\label{ij}\\
tr{\mathcal{K}}&=&\frac{2}{r}\label{trk}.
\end{eqnarray}
Thus, all the nontrivial components of the energy-momentum tensor are
\begin{eqnarray}
T_{tt}&=&-\frac{2m^2\alpha_1}{r}U,\label{ttt}\\
T_{tr}&=&\frac{2m^2\alpha_1}{r},\label{ttr}\\
T_{ij}&=&m^2\alpha_1 r\delta_{ij}\label{tij}.
\end{eqnarray}
Using the relations of the above energy-momentum tensor, it is easy to find that
\begin{equation}
T_{t\alpha}n^{\alpha}=0.
\end{equation}
Then, we straightforwardly obtain an important result from the background metric,
\begin{equation}
T_{b\alpha}n^{\alpha}=0
\end{equation}
where $b=\tau,i$.
Finally, making use of the equation of motion (\ref{EOM}),
the quantity ${B^k}_j$ reduces to the following form:
\begin{equation}
{B^k}_j=-(\frac{2m^2\alpha_1}{r_c}+3){\delta^k}_j.
\end{equation}
\subsection{The energy-momentum tensor in the general massive gravity}
Similarly, we will explicitly give out the detailed calculations
for the physical quantities $T_{\mu\nu}$ and ${B^k}_j$ in the general pure massive gravity.
Now let us start from the energy momentum tensor (\ref{emt})
for this model. Combining all nonzero components of this
matric ${\mathcal{K}^i}_j$, namely (\ref{sij}), with Eqs. (\ref{ij})
and (\ref{trk}), we can derive all the nontrivial components
of the energy-momentum tensor,
\begin{eqnarray}
T_{tt}&=&-U[m^2(\frac{2\alpha_1}{r}+\frac{2\alpha_2}{r^2})],\\
T_{tr}&=&m^2(\frac{2\alpha_1}{r}+\frac{2\alpha_2}{r^2}),\\
T_{ij}&=&m^2\alpha_1r\delta_{ij}.
\end{eqnarray}
 Furthermore, we can read out
\begin{equation}
{B^k}_j=[-3-m^2(\frac{2\alpha_1}{r_c}+\frac{2\alpha_2}{r^2_c})]{\delta^k}_j,
\end{equation}
as well as
\begin{equation}
T_{b\alpha}n^{\alpha}=0.
\end{equation}
\subsection{The high order estimation for the effective energy-momentum tensor}
Since the Navier-Stokes equation is produced by the momentum constraint
in (\ref{mc2}) at the order of $\lambda^1$, we must work out $T_{b\mu}n^{\mu}$
defined on near horizon boundary up to the order of $\lambda^1$. Therefore, taking
the perturbation behavior of metric into account, we can formally
express the total metric as
\begin{equation}
g^{(t)}_{\mu\nu}=g_{\mu\nu}+h_{\mu\nu},
\end{equation}
where $g_{\mu\nu}$ denotes the solution of the background spacetime
above, and $h_{\mu\nu}$ the perturbation part of metric in bulk.
In the holographic realization of the fluid/gravity duality,
one usually requires that the gravitational perturbations
on boundary do not change the behavior of the induced metric.
According to this requirement, we find that $h_{tr}(r_c)$ and $h_{ri}(r_c)$
on the near horizon boundary can be present, while other components
are vanishing. Note that we do not turn on the component of $h_{rr}$.
Then, we perform the near horizon expansion for these perturbational components
in power of $r_c-r_h$, which turn out to be $h_{tr}\sim O(r_c-r_h)$, $h_{ri}\sim O(r_c-r_h)$.
After straightforward calculation, the quantity
$T_{b\mu}n^{\mu}$ for both models turns out to be
\begin{eqnarray}
T_{\tau\nu}n^{\nu}=0+ O(\lambda^1)\\
T_{i\nu}n^{\nu}=0+O(\lambda^3).
\end{eqnarray}


\begin{thebibliography}{99}
\bibitem{Maldacena:1997re}
  J.~M.~Maldacena,
  Int.\ J.\ Theor.\ Phys.\  {\bf 38}, 1113 (1999)
  [Adv.\ Theor.\ Math.\ Phys.\  {\bf 2}, 231 (1998)]
  [hep-th/9711200].

\bibitem{Gubser:1998bc}
  S.~S.~Gubser, I.~R.~Klebanov and A.~M.~Polyakov,
  Phys.\ Lett.\ B {\bf 428}, 105 (1998)
  [hep-th/9802109].

\bibitem{Witten:1998qj}
  E.~Witten,
  Adv.\ Theor.\ Math.\ Phys.\  {\bf 2}, 253 (1998)
  [hep-th/9802150].

\bibitem{Aharony:1999ti}
  O.~Aharony, S.~S.~Gubser, J.~M.~Maldacena, H.~Ooguri and Y.~Oz,
  Phys.\ Rept.\  {\bf 323}, 183 (2000)
  [hep-th/9905111].

\bibitem{Damour1979}
T. Damour, (1979), Quelques propri∩et∩es m∩ecaniques,
∩electromagn∩etiques, thermodynamiques et quantiques des trous
noirs, Th`ese de doctorat d＊∩Etat, Uni- versit∩e Paris 6.
(available at http://www.ihes.fr/$\sim$ damour/Articles/). T.
Damour, (1982), Surface effects in black hole physics, in
Proceedings of the Second Marcel Grossmann Meeting on General
Relativity, Ed. R. Ruffini, North Holland , p. 587.

\bibitem{P-T}R.H. Price and K.S. Thorne, 
Phys. Rev. D 33, 915 (1986).

\bibitem{Jacobson:1995ab}
  T.~Jacobson,
  Phys.\ Rev.\ Lett.\   75, 1260 (1995) [arXiv:gr-qc/9504004].

\bibitem{PSS}G. Policastro, D.T. Son and A.O. Starinets, 
Phys. Rev. Lett. 87, 081601 (2001) {[}arXiv:hep-th/0104066{]};
JHEP 0209, 043 (2002) {[}arXiv:hep-th/0205052{]}.

\bibitem{KSS}P. Kovtun, D.T. Son and A.O. Starinets, 
JHEP 0310, 064 (2003) {[}arXiv:hep-th/0309213{]}.

\bibitem{B-L}A. Buchel and J.T. Liu, 
Phys. Rev. Lett. 93, 090602 (2004) {[}arXiv:hep-th/0311175{]}.

\bibitem{I-L}N. Iqbal and H. Liu, 
Phys. Rev. D 79, 025023 (2009) {[}arXiv:0809.3808{]}.

\bibitem{Bhattacharyya:2008kq}
   S.~Bhattacharyya, S.~Minwalla and S.~R.~Wadia,
  JHEP 0908, 059 (2009)
  [arXiv:0810.1545].

\bibitem{EFO} C. Eling, I. Fouxon and Y. Oz, 
 Phys. Lett. B 680, 496 (2009) {[}arXiv:0905.3638{]}.

\bibitem{Padmanabhan10rp}
  T.~Padmanabhan,
  Phys. Rev. D 83, 044048 (2011) [arXiv:1012.0119].

\bibitem{Wilsonian}
I. Bredberg, C. Keeler, V. Lysov and A. Strominger, 
JHEP 1103, 141 (2011) [arXiv:1006.1902].

\bibitem{Heemskerk10hk}
  I.~Heemskerk, J.~Polchinski,
  JHEP 1106, 031 (2011) [arXiv:1010.1264].

\bibitem{Faulkner10jy}
  T.~Faulkner, H.~Liu, M.~Rangamani,
  JHEP 1108, 051 (2011) [arXiv:1010.4036].

\bibitem{Bredberg:2011jq}
  I.~Bredberg, C.~Keeler, V.~Lysov and A.~Strominger,
  JHEP {\bf 1207}, 146 (2012)
  [arXiv:1101.2451 [hep-th]].

\bibitem{Compere:2011dx}
  G.~Compere, P.~McFadden, K.~Skenderis and M.~Taylor,
  JHEP {\bf 1107}, 050 (2011)
  [arXiv:1103.3022 [hep-th]].

\bibitem{Cai}R.-G. Cai, L. Li and Y.-L. Zhang, 
JHEP 1107, 027 (2011) [arXiv:1104.3281].

\bibitem{Bredberg:2011xw}
  I.~Bredberg and A.~Strominger,
  JHEP {\bf 1205}, 043 (2012)
  [arXiv:1106.3084 [hep-th]].

\bibitem{Niu:2011gu}
  C.~Niu, Y.~Tian, X.~N.~Wu and Y.~Ling,
  Phys.\ Lett.\ B {\bf 711}, 411 (2012)
  [arXiv:1107.1430 [hep-th]].

\bibitem{Matsuo11fk}
  S.~-J.~Sin, Y.~Zhou,
  JHEP 1105, 030 (2011) [arXiv:1102.4477];
  Y.~Matsuo, S.~J.~Sin and Y.~Zhou,
  JHEP {\bf 1201}, 130 (2012)
  [arXiv:1109.2698 [hep-th]].

\bibitem{Lysov11xx}
  V.~Lysov and A.~Strominger,
  ``From Petrov-Einstein to Navier-Stokes,''
  arXiv:1104.5502.

\bibitem{HL}
T. Huang, Y. Ling, W. Pan, Y. Tian and X. Wu, 
JHEP 1110, 079 (2011) [arXiv:1107.1464].

\bibitem{Huang:2011kj}
  T.~Z.~Huang, Y.~Ling, W.~J.~Pan, Y.~Tian and X.~N.~Wu,
  Phys.\ Rev.\ D {\bf 85}, 123531 (2012)
  [arXiv:1111.1576 [hep-th]].

\bibitem{Zhang:2012uy}
  C.~Y.~Zhang, Y.~Ling, C.~Niu, Y.~Tian and X.~N.~Wu,
  Phys.\ Rev.\ D {\bf 86}, 084043 (2012)
  [arXiv:1204.0959 [hep-th]].

\bibitem{Wu:2013kqa}
  X.~Wu, Y.~Ling, Y.~Tian and C.~Zhang,
  Class.\ Quant.\ Grav.\  {\bf 30}, 145012 (2013)
  [arXiv:1303.3736 [hep-th]].

\bibitem{Wu:2013mda}
  B.~Wu and L.~Zhao,
  Nucl.\ Phys.\ B {\bf 874}, 177 (2013)
  [arXiv:1303.4475 [hep-th]].

\bibitem{Ling:2013kua}
  Y.~Ling, C.~Niu, Y.~Tian, X.~N.~Wu and W.~Zhang,
  Phys.\ Rev.\ D {\bf 90}, no. 4, 043525 (2014)
  [arXiv:1306.5633 [gr-qc]].

\bibitem{Cai:2013uye}
  R.~G.~Cai, L.~Li, Q.~Yang and Y.~L.~Zhang,
  JHEP {\bf 1304}, 118 (2013)
  [arXiv:1302.2016 [hep-th]].

\bibitem{Cai:2014ywa}
  R.~G.~Cai, Q.~Yang and Y.~L.~Zhang,
  Phys.\ Rev.\ D {\bf 90}, no. 4, 041901 (2014)
  [arXiv:1401.7792 [hep-th]].

\bibitem{Cai:2014sua}
  R.~G.~Cai, Q.~Yang and Y.~L.~Zhang,
  JHEP {\bf 1412}, 147 (2014)
  [arXiv:1408.6488 [hep-th]].

\bibitem{Hao:2014xva}
  X.~Hao, B.~Wu and L.~Zhao,
  JHEP {\bf 1502}, 030 (2015)
  [arXiv:1412.8144 [hep-th]].

\bibitem{Hao:2015zxa}
  X.~Hao, B.~Wu and L.~Zhao,
  arXiv:1501.05146 [hep-th].

\bibitem{Pan:2015kia}
  W.~J.~Pan, Y.~Tian and X.~N.~Wu,
  Phys.\ Lett.\ B {\bf 752}, 1 (2016)
  [arXiv:1508.04972 [hep-th]].


\bibitem{Hao:2015pal}
  X.~Hao, B.~Wu and L.~Zhao,
  arXiv:1511.08281 [gr-qc].

\bibitem{Wu:2015pzg}
  C.~J.~Chou, X.~Wu, Y.~Yang and P.~H.~Yuan,
  Phys.\ Lett.\ B {\bf 761}, 131 (2016)
  [arXiv:1511.08691 [hep-th]].

\bibitem{Chou:2016qij}
  C.~J.~Chou, X.~Wu, Y.~Yang and P.~H.~Yuan,
  arXiv:1601.01946 [hep-th].

\bibitem{Fujisawa:2015riu}
  I.~Fujisawa and R.~Nakayama,
  arXiv:1511.08002 [hep-th].

\bibitem{Rebhan:2011vd}
  A.~Rebhan and D.~Steineder,
  Phys.\ Rev.\ Lett.\  {\bf 108}, 021601 (2012)
  [arXiv:1110.6825 [hep-th]].

\bibitem{Cheng:2014sxa}
  L.~Cheng, X.~H.~Ge and S.~J.~Sin,
  Phys.\ Lett.\ B {\bf 734}, 116 (2014)
  [arXiv:1404.1994 [hep-th]].

\bibitem{Cheng:2014qia}
  L.~Cheng, X.~H.~Ge and S.~J.~Sin,
  JHEP {\bf 1407}, 083 (2014)
  [arXiv:1404.5027 [hep-th]].

\bibitem{Ge:2014aza}
  X.~H.~Ge, Y.~Ling, C.~Niu and S.~J.~Sin,
  Phys.\ Rev.\ D {\bf 92}, no. 10, 106005 (2015)
  [arXiv:1412.8346 [hep-th]].

\bibitem{Jain:2014vka}
  S.~Jain, N.~Kundu, K.~Sen, A.~Sinha and S.~P.~Trivedi,
  JHEP {\bf 1501}, 005 (2015)
  [arXiv:1406.4874 [hep-th]].

\bibitem{Critelli:2014kra}
 {R.~Critelli, S.~I.~Finazzo, M.~Zaniboni and J.~Noronha,
  Phys.\ Rev.\ D {\bf 90}, 066006 (2014)
  [arXiv:1406.6019 [hep-th]].}

\bibitem{Jain:2015txa}
  S.~Jain, R.~Samanta and S.~P.~Trivedi,
  JHEP {\bf 1510}, 028 (2015)
  [arXiv:1506.01899 [hep-th]].

\bibitem{Bu:2014sia}
  Y.~Bu and M.~Lublinsky,
  Phys.\ Rev.\ D {\bf 90}, no. 8, 086003 (2014)
  [arXiv:1406.7222 [hep-th]].

\bibitem{Bu:2014ena}
  Y.~Bu and M.~Lublinsky,
  JHEP {\bf 1411}, 064 (2014)
  [arXiv:1409.3095 [hep-th]].

\bibitem{Donos:2015gia}
  A.~Donos and J.~P.~Gauntlett,
  Phys.\ Rev.\ D {\bf 92}, no. 12, 121901 (2015)
  [arXiv:1506.01360 [hep-th]].

\bibitem{Banks:2015wha}
  E.~Banks, A.~Donos and J.~P.~Gauntlett,
  JHEP {\bf 1510}, 103 (2015)
  [arXiv:1507.00234 [hep-th]].

\bibitem{Bu:2015ika}
  Y.~Bu and M.~Lublinsky,
  JHEP {\bf 1504}, 136 (2015)
  [arXiv:1502.08044 [hep-th]].

\bibitem{Bu:2015bwa}
  Y.~Bu, M.~Lublinsky and A.~Sharon,
  JHEP {\bf 1506}, 162 (2015)
  [arXiv:1504.01370 [hep-th]].

\bibitem{Bu:2015ame}
  Y.~Bu, M.~Lublinsky and A.~Sharon,
  JHEP {\bf 1604}, 136 (2016)
  [arXiv:1511.08789 [hep-th]].

\bibitem{Blake:2015epa}
  M.~Blake,
  JHEP {\bf 1509}, 010 (2015)
  doi:10.1007/JHEP09(2015)010
  [arXiv:1505.06992 [hep-th]].

\bibitem{Blake:2015hxa}
  M.~Blake,
  JHEP {\bf 1510}, 078 (2015)
  doi:10.1007/JHEP10(2015)078
  [arXiv:1507.04870 [hep-th]].


\bibitem{Vegh:2013sk}
  D.~Vegh,
  arXiv:1301.0537 [hep-th].

\bibitem{Blake:2013bqa}
  M.~Blake and D.~Tong,
  Phys.\ Rev.\ D {\bf 88}, no. 10, 106004 (2013)
  [arXiv:1308.4970 [hep-th]].

\bibitem{Andrade:2013gsa}
  T.~Andrade and B.~Withers,
  JHEP {\bf 1405}, 101 (2014)
  doi:10.1007/JHEP05(2014)101
  [arXiv:1311.5157 [hep-th]].

\bibitem{Amoretti:2014mma}
  A.~Amoretti, A.~Braggio, N.~Maggiore, N.~Magnoli and D.~Musso,
  Phys.\ Rev.\ D {\bf 91}, no. 2, 025002 (2015)
  [arXiv:1407.0306 [hep-th]].

\bibitem{Davison:2013jba}
  R.~A.~Davison,
  Phys.\ Rev.\ D {\bf 88}, 086003 (2013)
  [arXiv:1306.5792 [hep-th]].


\bibitem{Zeng:2014uoa}
  H.~B.~Zeng and J.~P.~Wu,
  Phys.\ Rev.\ D {\bf 90}, no. 4, 046001 (2014)
  [arXiv:1404.5321 [hep-th]].

\bibitem{Baggioli:2014roa}
  M.~Baggioli and O.~Pujolas,
  Phys.\ Rev.\ Lett.\  {\bf 114}, no. 25, 251602 (2015)
  [arXiv:1411.1003 [hep-th]].

\bibitem{Zhou:2015dha}
  Z.~Zhou, J.~P.~Wu and Y.~Ling,
  JHEP {\bf 1508}, 067 (2015)
  [arXiv:1504.00535 [hep-th]].

\bibitem{Sadeghi:2015vaa}
  M.~Sadeghi and S.~Parvizi,
  Class.\ Quant.\ Grav.\  {\bf 33}, no. 3, 035005 (2016)
  [arXiv:1507.07183 [hep-th]].

\bibitem{Alberte:2016xja}
  L.~Alberte, M.~Baggioli and O.~Pujolas,
  arXiv:1601.03384 [hep-th].

\bibitem{Hartnoll:2016tri}
  S.~A.~Hartnoll, D.~M.~Ramirez and J.~E.~Santos,
  JHEP {\bf 1603}, 170 (2016)
  [arXiv:1601.02757 [hep-th]].

\bibitem{deRham:2010kj}
  C.~de Rham, G.~Gabadadze and A.~J.~Tolley,
  Phys.\ Rev.\ Lett.\  {\bf 106}, 231101 (2011)
  doi:10.1103/PhysRevLett.106.231101
  [arXiv:1011.1232 [hep-th]].

\bibitem{Kovtun:2004de}
  P.~K.~Kovtun, D.~T.~Son and A.~O.~Starinets,
  Phys.\ Rev.\ Lett.\  {\bf 94}, 111601 (2005)
  [hep-th/0405231].

\bibitem{Brown:1992br}
  J.~D.~Brown and J.~W.~York, Jr.,
  Phys.\ Rev.\ D {\bf 47}, 1407 (1993)
  [gr-qc/9209012].

\bibitem{Matsuo:2009pg}
  Y.~Matsuo, T.~Tsukioka and C.~M.~Yoo,
  Europhys.\ Lett.\  {\bf 89}, 60001 (2010)
  [arXiv:0907.4272 [hep-th]].






\end{thebibliography}
\end{document}